\documentclass[onecolumn,groupedaddress]{revtex4}
\usepackage{graphicx}
\usepackage{epstopdf}
\usepackage{latexsym}
\usepackage{amssymb}
\usepackage{amsfonts} 
\usepackage{txfonts}
\usepackage{pxfonts}
\usepackage{color}
\usepackage[usenames,dvipsnames]{xcolor}
\usepackage{pdfpages}
\usepackage{tikz}
\usepackage{bbding}

\DeclareRobustCommand\solid  {\tikz[baseline=-0.6ex]\draw[thick] (0,0)--(0.5,0);}
\DeclareRobustCommand\dotted{\tikz[baseline=-0.6ex]\draw[thick,dotted] (0,0)--(0.54,0);}
\DeclareRobustCommand\dashed{\tikz[baseline=-0.6ex]\draw[thick,dashed] (0,0)--(0.54,0);}

\definecolor{newgreen}{rgb}{0.0, 0.5, 0.0}

\begin{document}

\title[An \textsf{achemso} demo]
  {Attractive Interaction between Fully Charged Lipid Bilayers in a Strongly-Confined Geometry}

\author{Tetiana Mukhina}
\affiliation{UPR 22/CNRS, Institut Charles Sadron, Universit\'e de Strasbourg, 23 rue du Loess, BP 84047 67034 Strasbourg Cedex 2, France}
\author{Arnaud Hemmerle}
\affiliation{UPR 22/CNRS, Institut Charles Sadron, Universit\'e de Strasbourg, 23 rue du Loess, BP 84047 67034 Strasbourg Cedex 2, France}
\author{Valeria Rondelli}
\affiliation{Dipartimento di Biotecnologie Mediche e Medicina Traslazionale, Universit\'a degli Studi di Milano, LITA, Via F.lli Cervi 93, 20090 Segrate, Italy}
\author{Yuri Gerelli}
\affiliation{Institut Laue-Langevin, 71 av. des Martyrs, BP 156, 38042 Grenoble Cedex, France}
\author{Giovanna Fragneto}
\affiliation{Institut Laue-Langevin, 71 av. des Martyrs, BP 156, 38042 Grenoble Cedex, France}
\author{Jean Daillant}
\affiliation{Synchrotron SOLEIL, L'Orme des Merisiers, Saint-Aubin, BP 48, F-91192 Gif-sur-Yvette Cedex, France}
\author{Thierry Charitat}
\email{thierry.charitat@ics-cnrs.unistra.fr}
\affiliation{UPR 22/CNRS, Institut Charles Sadron, Universit\'e de Strasbourg, 23 rue du Loess, BP 84047 67034 Strasbourg Cedex 2, France}

\begin{abstract}
We investigate the interaction between highly charged lipid bilayers in the presence of monovalent counterions. Neutron and X-ray reflectivity experiments show that the water layer between like-charged bilayers is thinner than for zwitterionic lipids, demonstrating the existence of counterintuitive electrostatic attractive interaction between bilayers. Such attraction can be explained by taking into account the correlations between counterions within the Strong Coupling limit, which falls beyond the classical Poisson-Boltzmann theory of electrostatics. Our results show the limit of the Strong Coupling continuous theory in a highly confined geometry and are in agreement with a decrease in the water dielectric constant due to a surface charge-induced orientation of water molecules.
\end{abstract}

\maketitle





Understanding electrostatic interactions between charged confined surfaces across aqueous electrolytes is important in many fundamental and applied research areas. For example, these interactions are crucial for controlling properties of colloidal suspension \cite{Norrish(nature1954),Belloni(jphyscondmat2000)}. In biology, many specific functions of cell membranes strongly depend on electrostatic interactions, such as interactions with biomolecules, membrane adhesion and cell-cell interactions \cite{andelman(revue)}. Electrostatic interactions between charged surfaces in water (dielectric permittivity $\epsilon_{\rm w}$) have been widely investigated both theoretically and experimentally in the case of two infinitely large planar walls with uniform surface charge density $\sigma_{s}$ and positively charged counterions of charge $qe$, where $q$ is the counterion valence, at temperature $T$. The Bjerrum length $\ell_{\rm B} = e^2/4\pi\epsilon_{\rm w} k_{\rm B} T \sim 0.7$ nm compares the electrostatic interaction between counter-ions to thermal energy. The Gouy-Chapman length $b=1/(2 \pi q \ell_{\rm B} \sigma_{s})$ characterizes the thickness of the diffuse counterion layer close to the membrane, without added salt, as a function of the charge density $\sigma_{s}$. The coupling constant $\Xi = \ell_{\rm B} q^2 /b =  2 \pi q^{3} \ell_{\rm B}^{2} \sigma_{s}$ quantifies the competition between the counterion-counterion interaction and thermal agitation $k_{\rm B}T$. For low coupling constant values, i.e. $\Xi \ll 1$, ion correlations are negligible and the ions distribution can be evaluated by the Poisson-Boltzmann (PB) theory in the mean-field approximation \cite{andelman(revue)}. PB theory predicts a repulsive pressure between similarly charged surfaces, and has been confirmed by numerous experiments  (for a review see \cite{Belloni(jphyscondmat2000)}). For high coupling constant values, i.e. $\Xi \gg 1$, the PB theory fails and correlations between ions start to be non-negligible. Indeed, strong-coupling (SC) theory and numerical simulations were developed to describe electrostatic interactions in this regime \cite{rouzina(jphyschem1996),Netz(epje2001),moreira(PRL2001),Moreira2002, Naji(PhysicaA2005),Samaj(PRL2011),trizac(softmatter2018)}, showing that identically charged plates can attract each other for large coupling parameters $\Xi \geq 20$.  Most of experimental investigations of SC limit have been carried out using divalent counterions to increase the coupling constant $\Xi$. In such conditions, an attraction, in good agreement with the strong coupling limit, was observed between mica surfaces \cite{kekicheff(Jchemphys1993)}, lamellar systems \cite{khan(jphyscem1985)} and between two charged vesicles \cite{salditt(2018)}. Recently, surface force apparatus experiments were performed to measure the compressibility modulus of charged membranes in presence of monovalent counterions. For large water separation distances ($d > 5$ nm), where effects related to the structure of water are negligible, a good agreement with weak coupling corrections, still in the repulsive regime ($\Xi\sim3$), was obtained \cite{herrmann(PRL2014)}. In this paper, we explore the poorly understood limit of strong confinement, where continuous theories reach their limits, as already highlighted numerically \cite{netz(langmuir2019)} and first evidenced by pioneering works on black films \cite{sonneville(langmuir2000)}.

\begin{figure}[h]
\begin{center}
\includegraphics[width=0.5\textwidth]{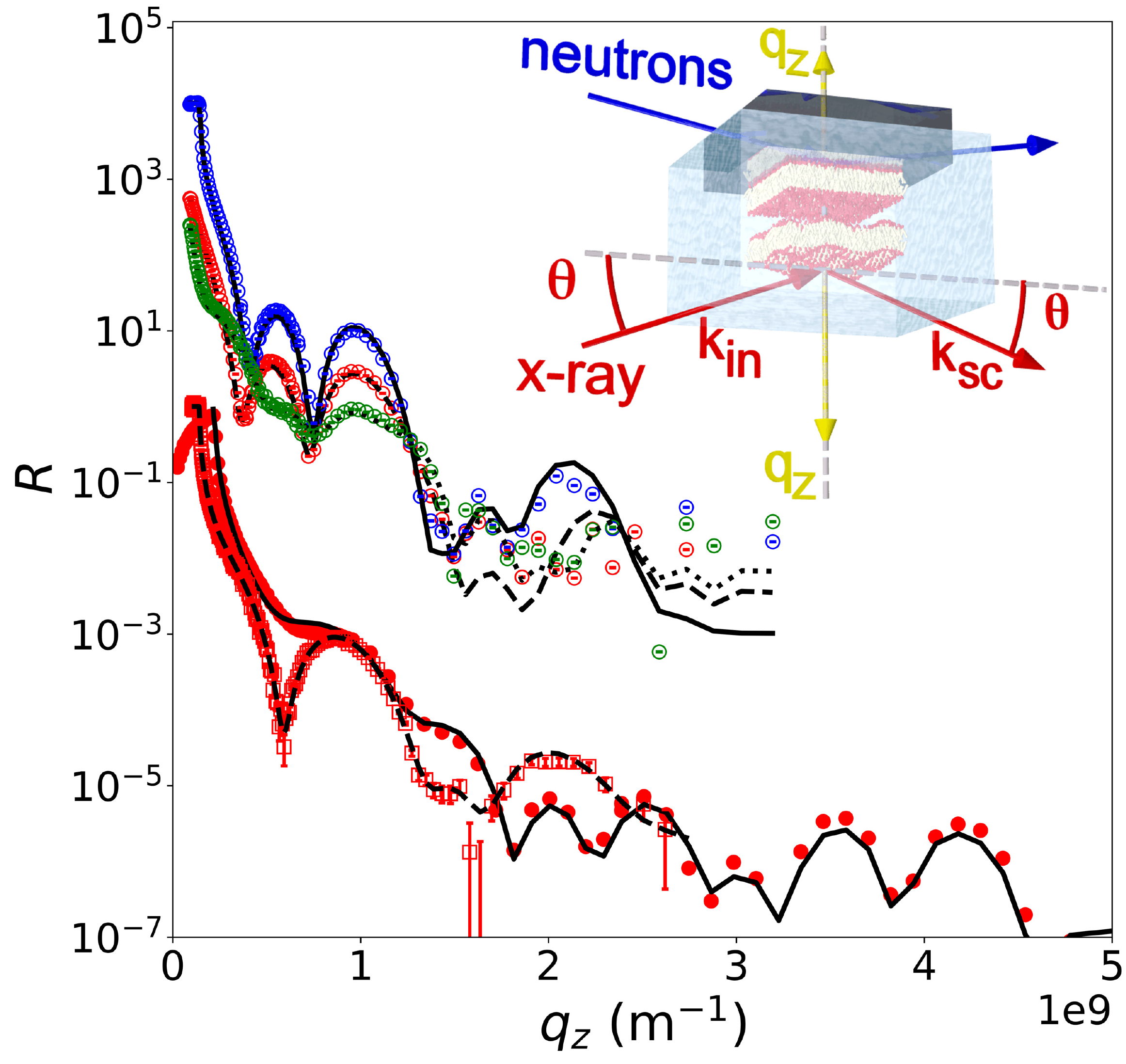}
\caption{(Bottom) Neutron reflectivity (NR, \textcolor{red}{$\square$}) and X-ray reflectivity (XRR, \textcolor{red}{$\bullet$}) for double DPPS bilayers at $T=40^\circ$C (in the gel phase). Dashed and solid lines are the best fits corresponding to the SLD profiles reported in Fig. \ref{figSLDED}. (Top) NR data in 3 different contrasts: (\textcolor{blue}{$\circ$}) in D$_2$O shifted of 4 decades for sake of clarity; (\textcolor{red}{$\circ$}) in SiMW and (\textcolor{newgreen}{$\circ$}) in H$_2$O for a DPPS triple bilayers at $T=25^\circ$C.}
\label{figref}
\end{center}
\end{figure}

Samples consisted in two supported bilayers deposited consecutively on ultra-flat silicon substrates \cite{charitat1999,daillant2005,Hemmerle(PNAS2012)} (see inset Fig. \ref{figref}). Highly charged double bilayers were prepared using DPPS (1,2-dipalmitoyl-sn-glycero-3-phospho-L-serine (sodium salt), Avanti Polar Lipids, Lancaster, Alabama, main transition temperature $T_{\rm m}=54^\circ$C). This system is denoted  DPPS$_2$-DPPS$_2$ in the following.  Zwitterionic double bilayers were prepared using DSPC (1,2-dipalmitoyl-sn-glycero-3-phosphocholine, Avanti Polar Lipids, Lancaster, Alabama, main transition temperature $T_{\rm m}=55^\circ$C) and the resulting system is denoted DSPC$_2$-DSPC$_2$. A double asymmetric bilayer, denoted DPPC/DPPS-DPPS/DPPC, and a triple DPPS bilayer, denoted DPPS$_2$-DPPS$_2$-DPPS$_2$, were also investigated. All the samples were prepared using the Langmuir-Blodgett (LB) and Langmuir-Schaefer (LS) deposition techniques (see more details in Supplementary Informations). 

\begin{figure}[h]
\begin{center}
\includegraphics[width=0.5\textwidth]{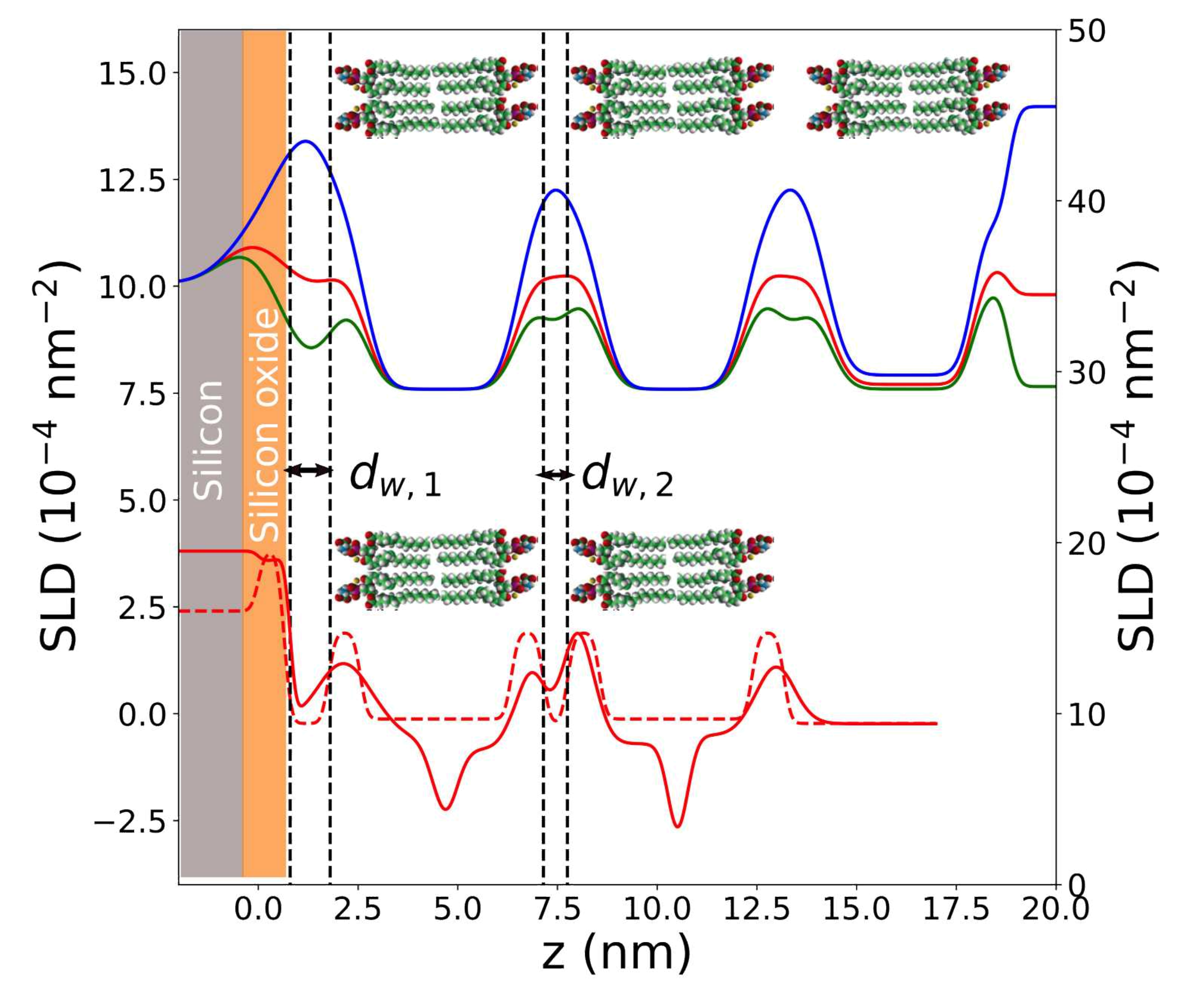}
\caption{SLD for NR (solid lines, left axis) and XRR (dashed lines, right axis) profiles corresponding to the best fits reported in Fig. \ref{figref}. (Bottom) DPPS$_2$-DPPS$_2$ at $T=40^\circ$C. (Top) DPPS$_2$-DPPS$_2$-DPPS$_2$ at $T=25^\circ$C in D$_2$O,  SiMW and H$_2$O contrast (shifted by 8.10$^{-4}$ nm$^{-2}$ for sake of clarity).}
\label{figSLDED}
\end{center}
\end{figure}

We have combined neutron reflectivity (NR) and X-ray reflectivity (XRR) to characterize with a high resolution ($\sim 0.1$ nm) the structure of double bilayers. NR measurements were performed on the D17 reflectometer \cite{Cubitt2002} at the Institut Laue-Langevin (ILL, Grenoble, France). 

The neutron beam was configured to illuminate through the silicon substrate the interface at which the sample was deposited. In order to apply the contrast variation method \cite{Crowley1991} to reduce ambiguities of the fits \cite{charitat1999,Koenig1996}, each system was measured against three different water solutions, namely 100\% H$_2$O (Scattering Length Density, SLD = $-0.56\cdot 10^{-6}$ \AA$^{-2}$), silicon-match water (SiMW, i.e. 62\% H$_2$O and 38\% D$_2$O by volume, SLD = $2.07\cdot 10^{-6}$ \AA$^{-2}$) and 100\% D$_2$O (SLD = $6.35\cdot 10^{-6}$ \AA$^{-2}$). XRR experiments were performed at the European Synchrotron Radiation Facility (ESRF, French CRG-IF, Grenoble, France) using a ${27}$ keV X-ray beam (wavelength ${\lambda=0.0459}$ nm). For both NR and XRR, specular reflectivity $R(q_z)$ is defined as the ratio between the specularly reflected and incoming intensities of a beam. $R(q_z)$ is expressed as a function of the wave vector transfer,  $q_z=4\pi/\lambda\sin\theta$ in the direction perpendicular to the sample surface, where $\theta$ is the grazing angle of incidence and reflection (see Fig. \ref{figref}). NR data was fitted with the AURORE software according to a discrete SLD profile \cite{Gerelli2016}, while XRR data was fitted using a hybrid gaussian continuous SLD profile (see \cite{daillant2005,malaquin10}). 

\begin{figure}[h!]
\centering
\includegraphics[width=.5\textwidth]{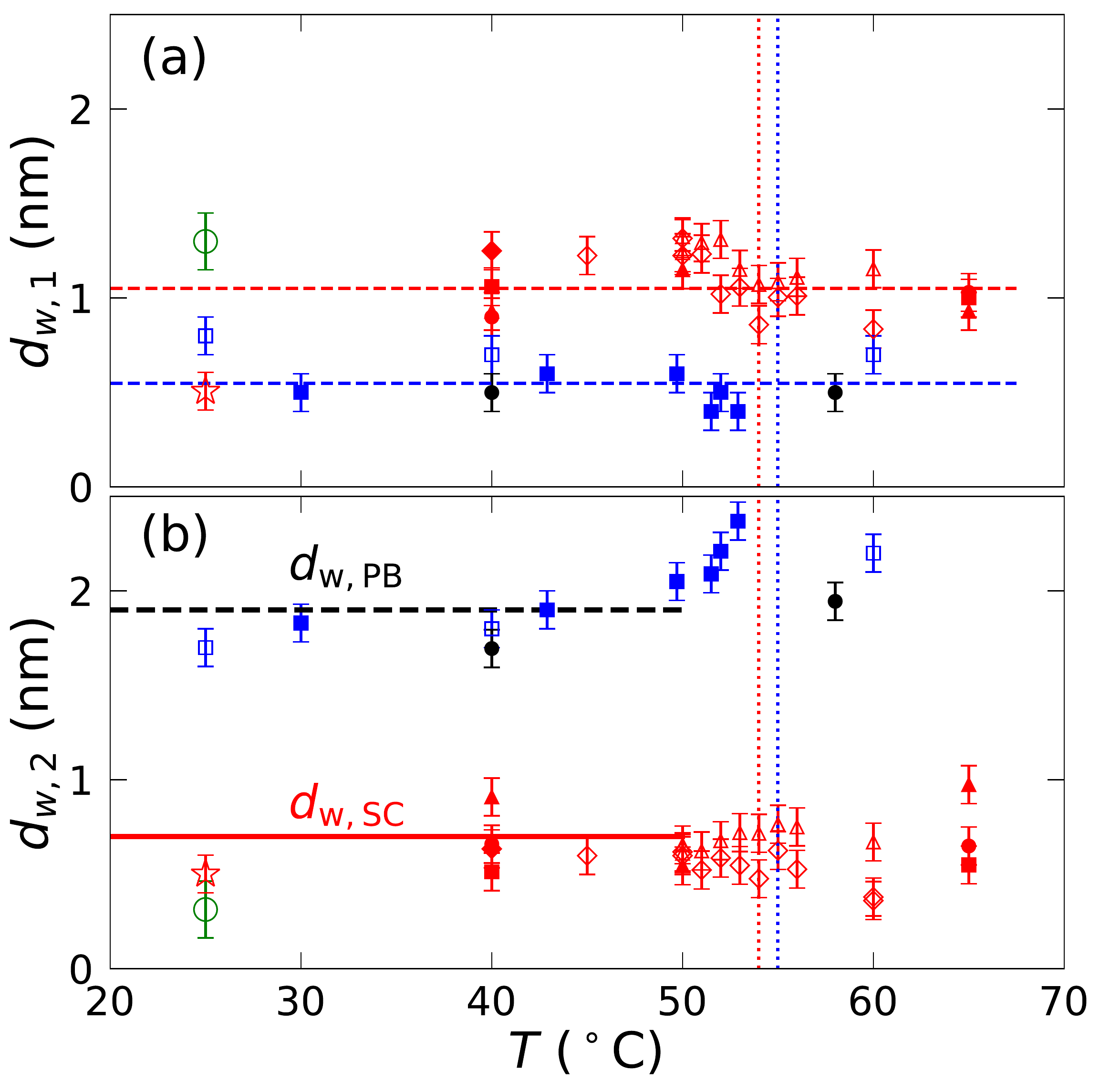}
\caption{(a) $d_{{\rm w},1}$ and (b) $d_{{\rm w},2}$ obtained from NR (open symbols) and XRR (closed symbols): DSPC$_2$-DSPC$_2$ with ($\bullet$) and without salt ($\textcolor{blue}{\blacksquare}$, $\textcolor{blue}{\Box}$); DPPS$_2$-DPPS$_2$ double bilayer (XRR, 4 different samples \textcolor{red}{$\blacksquare$}, \textcolor{red}{$\blacktriangle$}, \textcolor{red}{$\bullet$}, \textcolor{red}{$\blacklozenge$}) and (NR, 2 different samples \textcolor{red}{$\bigtriangleup$}, \textcolor{red}{$\lozenge$}); DPPS$_2$-DPPS$_2$-DPPS$_2$ (\textcolor{red}{$\bigcirc$}); DSPC/DPPS-DPPS/DSPC double asymmetric bilayer (\textcolor{red}{\FiveStarOpen}). (a) Dashed lines are guide for the eye corresponding to average values. (b) Black dashed line ($\dashed$) corresponds to $d_{w, vdW}=1.9$ nm and solid red line ($\textcolor{red}{\solid}$) corresponds to $d_{w,SC}=0.7$ nm. Dotted lines corresponds to the gel to fluid transition temperature for DSPC ($\textcolor{blue}{\dotted}$) and DPPS ($\textcolor{red}{\dotted}$) bilayers.}
\label{figdw1dw2vsT}
\end{figure}

\begin{figure}[h!]
\centering
\includegraphics[width=.5\textwidth]{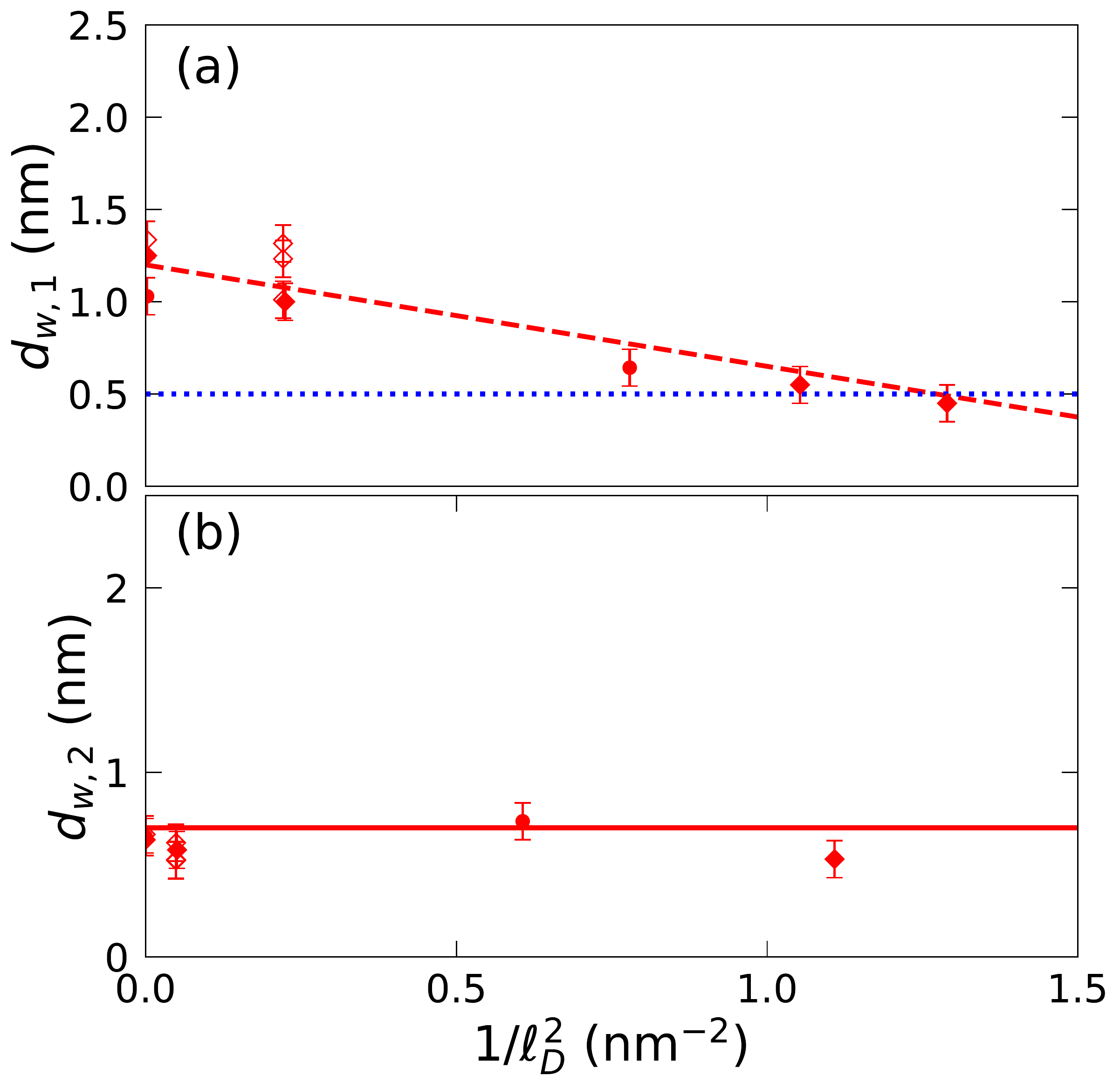}
\caption{(a) $d_{{\rm w},1}$ and (b) $d_{{\rm w},2}$ vs $1/\ell_D^2$ (same notations as Fig. \ref{figdw1dw2vsT}). (a) Blue dotted line ($\textcolor{blue}{\dotted}$) corresponds to $d_{w, 1}=0.5$ nm, average value for DSPC case and  ($\textcolor{red}{\dashed}$) to osmotic pressure effect; (b) Red solid line ($\textcolor{red}{\solid}$) corresponds to strong coupling theory $d_{w,SC}=0.7$ nm.}
\label{figdw1dw2vsld}
\end{figure}

Fig. \ref{figref} shows NR and XRR data for DPPS$_2$-DPPS$_2$ and the best fits corresponding to the SLD profiles are shown in Fig. \ref{figSLDED}. Both NR and XRR profiles are in good quantitative agreement. We define the water thicknesses between the substrate and the first bilayer ($d_{{\rm w},1}$) and between the two bilayers ($d_{{\rm w},2}$) by the distance between inflection points in the SLD profiles (see dashed line Fig. \ref{figSLDED}). We have investigated the variation of $d_{{\rm w},1}$, the thickness of the water layer between the silicon oxide and the bilayer, and $d_{{\rm w},2}$, the interbilayer water thickness, as a function of the temperature (see Fig. \ref{figdw1dw2vsT}(a) and (b)) and as a function of the Debye length $\ell_{\rm D}=\left(\epsilon_{\rm w} k_BT/2e^2 c\right)^{1/2}$ upon changes in NaCl concentration $c$ (0.01-0.3 M) (see Fig. \ref{figdw1dw2vsld}(a) and (b)) for DPPS$_2$-DPPS$_2$ and DSPC/DPPS-DPPS/DSPC samples. The obtained result was compared to our previous work on DSPC$_2$-DSPC$_2$ in absence and in presence of NaCl (0.5 M) \cite{Fragneto2003,Hemmerle(PNAS2012)}. 

As expected, $d_{{\rm w},1}$ is larger when the first bilayer is charged ($d_{{\rm w},1}\sim 1.0 \pm 0.1$ nm for DPPS$_2$-DPPS$_2$) than when it is composed of zwitterionic phospholipids ($d_{{\rm w},1}\sim 0.6 \pm 0.1$ nm for DSPC$_2$-DSPC$_2$). $d_{{\rm w},1}$ is mainly controlled by the weak electrostatic repulsive interaction between the first monolayer and the silicon oxide layer of the substrate that is negatively charged, because of the experimental conditions and the surface treatment prior to the deposition (see more details in Supplementary Informations). With the exception of a small variation around the transition temperature, within the resolution limit of the techniques, we did not observe a significant variation of $d_ {{\rm w}, 1}$ with temperature. This shows that the entropic contributions related to membrane fluctuations are negligible. Finally, we observed that $d_{{\rm w},1}$ decreases as the salt concentration increases (see Fig. \ref{figdw1dw2vsld}(a)), reaching values close to those observed for DSPC$_2$-DSPC$_2$ case ($d_{{\rm w},1}\sim 0.5$ nm). This is in good qualitative agreement with an osmotic pressure effect as described in \cite{Hishida(2017)} (see also Supplementary Information).

We now discuss the case of $d_{{\rm w},2}$ (see Fig. \ref{figdw1dw2vsT}(b)). As in the case of $d_{{\rm w},1}$, $d_{{\rm w},2}$ values obtained from NR and XRR data are in good agreement. In the case of DPPS$_2$-DPPS$_2$ charged systems (red symbols in Fig. \ref{figdw1dw2vsT}(b)), we observe that $d_{{\rm w},2}\simeq 0.6\pm 0.1$ nm is smaller than in the case of the zwitterionic DSPC$_2$-DSPC$_2$ systems, for which $d_{{\rm w},2}\simeq 1.8-2.5$ nm (blue and black symbols). Contrary to the case of zwitterionic lipids, we did not observe variations of $d_{{\rm w},2}$  with temperature, indicating that entropic contributions related to membrane fluctuations are negligible. The decrease of $d_{{\rm w},2}$ in the fully charged systems clearly indicates that the attractive contribution to the interaction potential is increased in the case of two highly like-charged membranes.  We exploited this result to deposit, by means of the LB technique, up to 5 DPPS monolayers (see more details in Supplementary Information), leading to a triple bilayer systems (DPPS$_2$-DPPS$_2$-DPPS$_2$, see NR data in Fig. \ref{figref}). The corresponding SLD profiles, shown in Fig. \ref{figSLDED}, demonstrate the high quality and structural integrity of all bilayers. If zwitterionic molecules such as phosphocholine are used, the deposition of more than three successive monolayers by LB is not possible. Try to deposit a fourth monolayer usually leads to the partial removal of the third one (see Supplementary Information). DPPS$_2$-DPPS$_2$-DPPS$_2$ was also used to quantify attractive interactions between bilayers that are far from the solid substrate. In this case, the interbilayer water thickness $d_{{\rm w},2}$ was found to be constant ($d_{{\rm w},2}=0.3\pm 0.2$ nm), demonstrating that the substrate has only a minor influence on the deposition after the first monolayer. The role of electrostatic interactions is confirmed by complementary experiments on asymmetric sample DSPC/DPPS-DPPS/DSPC for which we obtained $d_{{\rm w},2}\simeq 0.5\pm 0.1$ nm, a value in agreement with those obtained for the DPPS$_2$-DPPS$_2$ samples (see Fig. \ref{figdw1dw2vsT}(b)). Finally,  an increasing of NaCl concentration in the solution has a negligible effect on the value of $d_{{\rm w},2}$, as clearly evidenced in Fig. \ref{figdw1dw2vsld}(b).

To compare our experimental results with existing theoretical models, we have to take into account the different contributions to the interactions between adjacent bilayers. First, the short range hydration repulsion is described by a potential $U_{\rm hyd}=P_h\lambda_h \exp{\left(-d_{\rm w}/\lambda_h\right)}$ with a hydration pressure $P_h = 4\cdot10^9$ Pa and a decay length $\lambda_h = 0.1-0.2$ nm. We model the van der Waals attractive contribution as $U_{\rm vdW} = -H/12\pi\left(d_{\rm w}+2d_{\rm head}\right)$ with the Hamaker constant $H=5.3\cdot 10^{-21}$ J and a head thickness $d_{\rm head}\sim 0.5$ nm.  Finally, as described in our previous work \cite{Hemmerle(PNAS2012)}, the small amount of charges due to the amphoteric character of the phosphatidylcholine group ($\sigma\sim 0.001$ e/nm$^2$), leads to a weak electrostatic repulsion corresponding to the Ideal Gas limit of the mean-field PB theory $U_{\rm PB}\left(z\right)=-2k_{\rm B}T\sigma_S\log{z}$ \cite{andelman(revue)}. In gel phase, neglecting entropic contribution, the minimization of $U_{\rm vdW}\left(z\right) + U_{\rm hyd}\left(z\right) + U_{\rm PB}\left(z\right)$, leads to an equilibrium value for the water thickness $d_{w, {\rm PB}}\simeq 1.9$ nm (black dashed line in Fig. \ref{figdw1dw2vsT}(b)). Such a value is in good agreement with those obtained for a DSPC$_2$-DSPC$_2$ system in gel phase. As demonstrated in our previous work \cite{Hemmerle(PNAS2012)}, the increase of $d_{{\rm w},2}$ close to the gel-fluid transition shown in Fig. \ref{figdw1dw2vsT}(b) is well described by taking into account entropic contribution and electrostatic repulsion in the framework of PB theory.

\begin{figure}[h!]
\centering
\includegraphics[width=.5\textwidth]{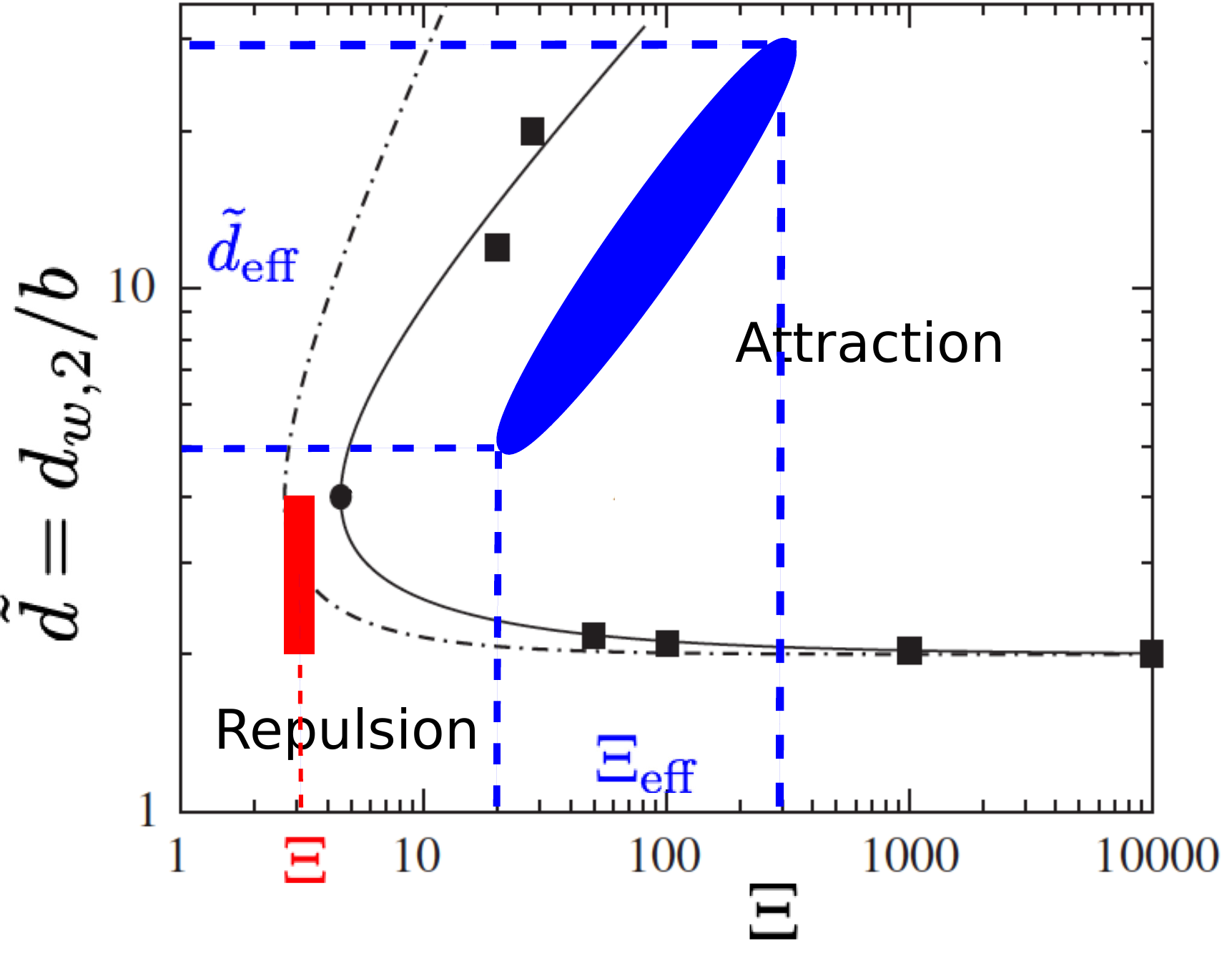}
\caption{Phase diagram showing attraction and repulsion regimes in terms of the rescaled water thickness $\tilde{d}=d_{{\rm w},2}/b$ and as a function of the coupling constant $\Xi$ (figure adapted from \cite{Samaj(PRL2011)}). The dashed-dotted line is the original Virial Strong Coupling theory from \cite{Netz(epje2001)} and black squares are the Monte Carlo simulation data from \cite{moreira(PRL2001)}. Solid line corresponds to Wigner Strong Coupling theory from \cite{Samaj(PRL2011)}. Our experiments corresponds to the red area using $\epsilon_{\rm w}=80\epsilon_0$ ($\Xi\sim 2.5-4, {\tilde d}\sim 2-4$) and by the blue one, taking into account renormalization induced by water orientation ($10 \leq \epsilon_{\rm w} \leq 30 \Leftrightarrow 20 \leq \Xi_{\rm eff} \leq 300$, $5 \leq \tilde{d}_{\rm eff} \leq 30$).}
\label{figTrizacNetz}
\end{figure}

For DPPS double and triple bilayers, the entropic contribution is clearly negligible, as we observe no significant variation of $d_{{\rm w},2}$ with the temperature. Regarding electrostatic interaction between charged surfaces, the area per molecule of DPPS can be estimated to be equal to 0.55 nm$^2$ \cite{thurmond(BiophysJ1991)}. At pH=5-6, taking into account the respective pKa of the phosphate, ammonium and carboxylate groups \cite{marsh2013handbook}, we obtain a charge density $\sigma\sim 0.8-1.5$ e/nm$^2$. By estimating  the coupling constant with $\epsilon_{\rm w}\simeq80\epsilon_0$ we obtain $\Xi \simeq 2.5-4$, which is outside the attractive zone described by the SC theory \cite{Netz(epje2001),Samaj(PRL2011)} (see red area on Fig. \ref{figTrizacNetz}). In this limit, using realistic values of $P_h$, $\lambda_h$ and $H$ it is not possible to access equilibrium values of $d_{{\rm w},2}$ smaller than 1.5 nm. In such strongly confined limit, the rotational degrees of freedom of water dipoles are expected to be frozen near surfaces, inducing a strong decrease of the dielectric permittivity. Fumagalli {\it et al.} \cite{Fumagalli1339} experimentally demonstrated the presence of an interfacial two to three molecules thick layer with $\epsilon_{\rm w}\sim 2\epsilon_0$, a value close to the limit for water at optical frequencies ($\epsilon_{\rm w}\sim1.8\epsilon_0$).  Since $\Xi\sim 1/\epsilon^2$, this leads to a strong increase of the coupling constant. Recently, using water-explicit numerical simulations of decanol bilayers with variable charge density, Schlaich {\it et al.} \cite{netz(langmuir2019)} demonstrated that an attractive behavior can appear at a moderate surface density ($\sim 0.77$ e/nm$^2$, $\Xi\sim 3$), a value close to the lowest estimation of the charge density in our experiments. Their numerical results are in good agreement with an effective coupling constant $\Xi_{\rm eff}\sim 20$, corresponding to a decrease of $\epsilon_{\rm w} $ to 30. For the upper limit of charge density ($\sim 1.5$ e/nm$^2$), we can extrapolate an effective dielectric constant of the order of 10. The blue area in Fig. \ref{figTrizacNetz} give the corresponding effective values of $\Xi_{\rm eff}$ ($10<\Xi_{\rm eff}<300$) and ($5<\tilde{d}_{\rm eff}<15$), taking into account the effective dielectric constant. It is clear that our experiments fall in the attractive regime, well described by SC theory where it is valid to use the analytical expression for the electrostatic pressure $P_{\rm SC} = 2\pi\ell_{\rm B}\sigma_S^2\left(2b/z-1\right)k_{\rm B} T$ \cite{Netz(epje2001),Naji(PhysicaA2005)}. The decrease of $\epsilon_{\rm w}$ also induced a decrease of the Hamaker constant by one order of magnitude \cite{parsegian2005van} and it is possible to neglect the van der Waals interactions compared to the SC term. By minimizing the potential $U\left(z\right)$ ~:
 \begin{equation}
U\left(z\right) =U_{\rm vdW}\left(z\right) + U_{\rm hyd}\left(z\right) - 2\pi\ell_{\rm B}b\sigma_S^2k_{\rm B} T\left(2\log{\left(\frac{z}{b}\right)}-\frac{z}{b}\right)
\label{USC}
 \end{equation}
 we obtain an equilibrium value of $d_{w,{\rm SC}}=0.7$ nm, in very good agreement with our experimental data as shown in Fig. \ref{figdw1dw2vsT}(b) and Fig. \ref{figdw1dw2vsld}(b).
 
Electrostatic interactions between highly charged double bilayers, in presence of monovalent counterions and in strong confinement, have been investigated by measuring the equilibrium distance between like-charged bilayers. In a consistent set of experimental data obtained on a model system close to an ideal theoretical configuration (planar geometry, negligible fluctuations), we have demonstrated the presence of an attractive electrostatic interaction, which is in contradiction to continuous theories. Our results are in agreement with recent water-explicit numerical simulations \cite{netz(langmuir2019)} predicting that ion correlations can have tremendous effects even for moderate surface charge densities and in the presence of monovalent counterions. The obtained results provide a deeper understanding of electrostatic interactions in strong confinement beyond the continuous models, where  the discrete nature of charges  must be taken into account explicitly.

\section*{Acknowledgement}

The authors thanks L. Malaquin and S. Micha for assistance during the experiments, P. K\'ekicheff, A. Johner, C. Loison and R. Netz for fruitful discussions. Awarded beamtime at the ILL (10.5291/ILL-DATA.EASY-341) and at the ESRF is gratefully acknowledged. Supports from the Labex NIE 11-LABX-0058-NIE (Investissement d'Avenir programme ANR-10- IDEX-0002-02) and PSCM facilities at the ILL for sample preparation are gratefully acknowledged. T. Mukhina thanks ILL for a PhD grant.

\section*{Supplementary Informations}

Supporting Information Available: Detailed description of sample preparation and data analysis. Detailed description of the osmotic pressure effect (PDF).


\newpage

\pagestyle{empty}


\section*{Supplementary material}

# Supplementary Information for Attractive Interaction between Fully Charged Lipid Bilayers in a Strongly-Confined Geometry

T. Mukhina[1,2], A. Hemmerle[1,3], V. Rondelli[4], Y. Gerelli[2], G. Fragneto[2], J. Daillant[3], T. Charitat[1]

November 6, 2019

[1]UPR 22/CNRS, Institut Charles Sadron, Université de Strasbourg, 23 rue du Loess, BP 84047 67034 Strasbourg Cedex 2, France
[2]Institut Laue-Langevin, 71 av. des Martyrs, BP 156, 38042 Grenoble Cedex, France
[3]Synchrotron SOLEIL, L'Orme des Merisiers, Saint-Aubin, BP 48, F-91192 Gif-sur-Yvette Cedex, France
[4]Dipartimento di Biotecnologie Mediche e Medicina Traslazionale, Universitá degli Studi di Milano, LITA, Via F.lli Cervi 93, 20090 Segrate, Italy

## 1 Materials

DPPS ($L-\alpha$ 1,2-dipalmitoyl-sn-glycero-3-phospho-L-serine (sodium salt)) and DSPC ($L-\alpha$ 1,2-distearoyl-sn-glycero-3-phosphocholine) were purchased from Avanti Polar Lipids (Lancaster, Alabama). Pure water (resistivity 18.2 M$\Omega$.cm) was obtained from a Millipore purification system. D$_2$O was supplied by the Institut Laue-Langevin (ILL, Grenoble, France). All the solvents used in the preparation were purchased from Sigma-Aldrich (Saint-Quentin Fallavier, France) and used without further purifications.

## 2 Sample preparation

Supported bilayers were prepared on 25 cm$^2$ surfaces of n-doped silicon single crystals cut along the (111) direction. The surface roughness was lower than 5 Å, as determined by neutron reflectometry (NR) and X-ray reflectometry (XRR) experiments. Prior to their use, silicon blocks were cleaned using ultra-sonic bath in the following sequence of solvents: acetone, chloroform, ethanol and water. The blocks were sonicated in each solvent for 20 minutes. They were then dried under a nitrogen stream and treated with air-plasma (Harrick Plasma, US) for 2 minutes (for NR measurements) or with UV-Ozone (BioForce Nanosciences, USA) for 20 minutes (for XRR measurements) to make their surface as hydrophilic as possible. Depositions of double bilayer systems were performed by Langmuir Blodgett (LB) and Langmuir Schaefer (LS) techniques using the 611 and 1212D NIMA Langmuir troughs (Coventry, UK) available at the Partnership for Soft Condensed Matter (PSCM, Grenoble, France). Monolayers at the air-water interface (prepared as described in [S1]) were compressed to a lateral pressure $\pi$ = 40 mN/m. While the pressure was kept constant, the monolayer was transferred to the silicon block by LB. In the case of double bilayer systems this procedure was repeated three times while for triple bilayer systems it was repeated five times. In all cases, the final monolayer was deposited by a LS transfer. The samples were sealed inside a solid-liquid cell to be mounted on the neutron or x-ray reflectometers. The temperature of the cell was controlled using a water circulation bath. The water reservoir of the cell was connected to a HPLC pump in order to be able to change the subphase for the application of the contrast variation method [S2].

## 3 LB transfer efficiency

During the LB deposition the transfer quality was evaluated by monitoring the transfer ratio, TR. This ratio is defined as

$$TR = w \times \frac{\Delta Y_{block}}{\Delta A_{mono}} \tag{S1}$$

where $\Delta Y_{block}$, $\Delta A_{mono}$ and $w$ are the substrate vertical displacement, the variation of the area occupied by the Langmuir monolayer and the substrate section full perimeter respectively. Fig. S1 shows the variation of $Y_{block}$ vs. $A_{mono}$ during the deposition of DSPC and DPPS monolayers.

We can observe that, contrarily to the case of DPPS, it was not possible to deposit more than 3 successive DSPC monolayers; the attempt of depositing a fourth monolayer on the top of the third one failed systematically. More importantly, the trend shown in Fig. S1b indicates that the deposition did not simply fail, but promoted the removal of the lipids composing the third monolayer, as indicated by the increase of $A_{mono}$. In fact, the monolayer area should decrease upon transfer or stay constant in the case of null transfer. For this reason, the last monolayer for zwitterionic systems was deposited using the LS technique, which proved to be more efficient. This problem was not present in the case of charged monolayers (DPPS) but we have decided to apply the same preparation protocol to all the samples. Therefore, the last DPPS monolayer was deposited by LS as well. However, it is important to note that the possibility of depositing more than 3 charged monolayers by LB is a direct result of the attractive interaction between likely-charged systems described in the main text .

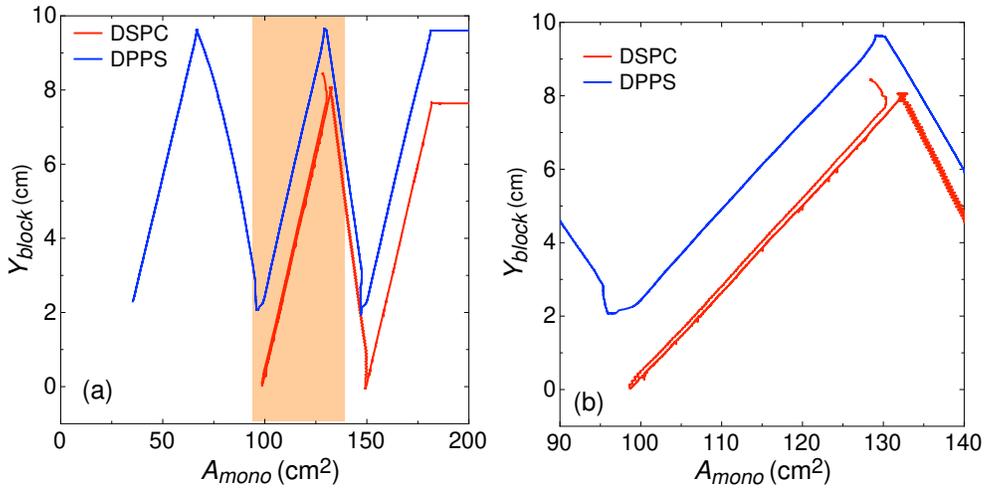

Figure S1: (a) Successive deposition of lipid monolayers on silicon block at $\Pi$ = 40 mN/m. For DSPC (red) the fourth deposition attempt removes the third layer and for DPPS (blue) successful five layer deposition was achieved. (b) Zoom on the highlighted zone of (a) (third and fourth deposition).

The TR values obtained from the analysis of $Y_{block} - A_{mono}$ curves for the samples presented in the main manuscript are reported in Fig. S2. These $TR$s are all close to 1, demonstrating the full coverage of the samples which is in good agreement with NR and XRR data.

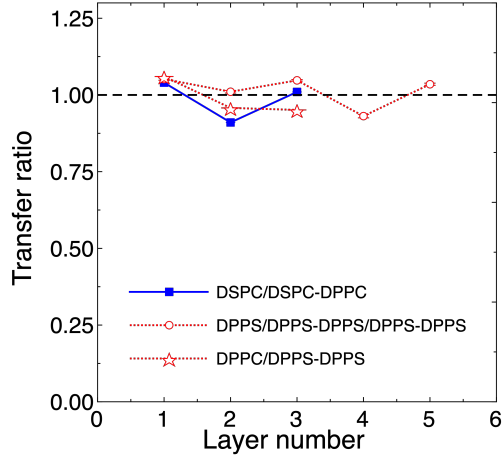

Figure S2: Transfer ratio for the successively deposited layers at $\Pi$ = 40 mN/m: (■) DSPC, (○) DPPS and (☆) DSPC/DPPS-DPPS.

## 4 Neutron and X-ray reflectometry

Specular reflectivity, $R(q)$, is defined as the ratio between the specularly reflected and incoming intensities of a beam, which is measured as a function of the wave vector transfer, $q = 4\pi/\lambda \sin\theta$ in the direction perpendicular to the reflecting surface, where $\theta$ is the angle of reflection and $\lambda$ the wavelength of the probe.

### 4.1 Neutron reflectometry

Neutron reflectivity (NR) measurements were performed on the D17 reflectometer [S3] at the Institut Laue-Langevin (ILL, Grenoble, France). The instrument was operated in time of flight mode with a wavelength range from 2 to 20 Å. Each measurement was performed at two reflection angles, $\theta_1 = 0.8°$ (resolution $\Delta q/q$ = 2.7%) and $\theta_2 = 3.2°$ (resolution $\Delta q/q$ varied linearly from 3.8 % to 13 %). This resulted in a $q_z$-range for specular reflectivity ranging from 0.005 to 0.3 Å$^{-1}$. The neutron beam was configured to illuminate the interface at which the sample was deposited through the silicon substrate. In order to apply the contrast variation method [S2] to reduce ambiguities of the fits [S2, S4, S1], each system was measured against three different water solutions, namely as 100% H$_2$O (Scattering Length Density, SLD = $-0.56 \cdot 10^{-6}$ Å$^{-2}$), silicon-match water (SiMW, i.e. 62% H$_2$O and 38% D$_2$O, SLD = $2.07 \cdot 10^{-6}$ Å$^{-2}$) and 100% D$_2$O (SLD = $6.35 \cdot 10^{-6}$ Å$^{-2}$).

### 4.2 X-ray reflectometry

X-ray reflectivity (XRR) experiments were performed at the CRG-IF beamline of the European Synchrotron Radiation Facility (ESRF, Grenoble, France) using a 27 keV x-ray beam. Measurements were performed at a fixed incident wavelength $\lambda$ = 0.459 Å with the incident angle varying from 0 ° to 1.05 °. The reflected intensity was recorded using a NaI (Tl) scintillator. The $q_z$-range for specular reflectivity ranged from 0.0 to 0.5 Å$^{-1}$. As it is visible in Fig. 1 in the main manuscript, footprint corrections were not applied. In fact, the instrumental configuration (slits, sample surface size, incidence angles) was

chosen to avoid over-illumination for the $q_z$-range beyond the critical edge (located at $q_c = 0.023 \text{Å}^{-1}$ for the water-silicon interface).

## 4.3 Data analysis

Because of the general phase problem typical for scattering experiments [S5], $R(q)$ curves need to be modeled to find the appropriate scattering length density (SLD) profiles. In order to reduce the number of free parameters during the analysis of double bilayer systems, the bare silicon substrates were characterized prior to the deposition by means of NR and XRR. The errors for the parameters were determined by evaluating the variation of the $\chi^2$ upon changes in parameter values.

### 4.3.1 SLD profile for NR

NR data was fitted with the AURORE software [S6], where the specular reflectivity is calculated by the Parratt's recursive formalism for stratified interfaces [S7]. NR curves originated from the same bilayer systems and collected in different contrasts were analysed simultaneously by applying the same structural model as described in [S6]. Only the subphase SLD was changed for different contrasts. Double bilayers were fitted using a 9 slabs model (see Fig. S3(a)). Slab $i$ is described by a scattering length density $SLD_i$, a fraction of solvent ($H_2O$ or $D_2O$) $f_{w,i}$ and a thickness $d_i$:

1. $i = SiO_2$, silicon oxide layer;
2. $i = w, 1$, first water layer;
3. $i = h11$, lipid head of the first monolayer;
4. $i = ch1$, lipid chain of the first bilayer;
5. $i = h12$, lipid head of the second monolayer;
6. $i = w, 2$, second water layer;
7. $i = h21$, lipid head of the third monolayer;
8. $i = ch2$, lipid chain of the second bilayer;
9. $i = h22$, lipid head of the fourth monolayer.

Each interface between two slabs $i$ and $j$ is also characterized by a width $\sigma_{i,j}$. Values obtained from best fits are given in Table S2 for a DPPS double bilayer and Table S3 for a triple bilayer.

### 4.3.2 SLD profile for XRR

XRR data was fitted using a 1G-hybrid gaussian continuous profile (see [S8, S9] and Fig. S3(b)).

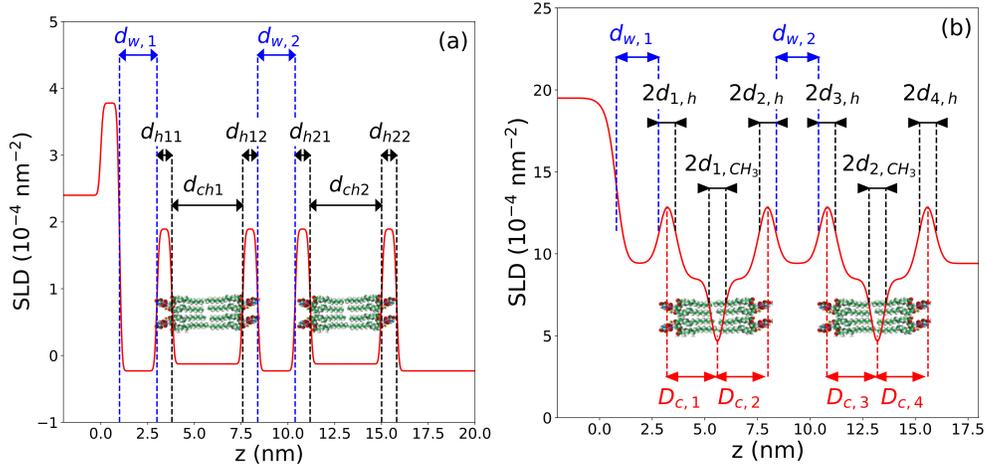

Figure S3: (a) Definition of the parameter used for SLD profile in case of NR. For sake of clarity we used non-realistic values for $\sigma_{i,j}$, $d_{w,1}$ and $d_{w,2}$ ($\sigma_{i,j}$ = 0.1 nm, $d_{w,1}$ = 2 nm, $d_{w,2}$ = 2 nm). (b) Definition of the parameter used for SLD profile in case of XR (1G-hybrid model). For sake of clarity we used non-realistic values for $d_{w,1}$ and $d_{w,2}$ ($\sigma_{i,j}$ = 0.1 nm, $d_{w,1}$ = 2 nm, $d_{w,2}$ = 2 nm)

### 4.3.3 Fitted parameters

Table S1: Serie A : Values of structural parameters obtained from the fit of the specular XRR data for the highly charged DPPS double bilayer.

| Parameter | T = 40°C |
|---|---|
| $\rho_{SiO_2}$ [$10^{-4}/nm^2$] | 19 ± 2 |
| $d_{SiO_2}$ [nm] | 0.8 ± 0.2 |
| $\sigma_{SiO_2}$ [nm] | 0.2 ± 0.1 |
| $d_{w,1}$ [nm] | 0.97 ± 0.1 |
| $\rho_{1,head}$ [$10^{-4}/nm^2$] | 14 ± 2 |
| $\rho_{2,head}$ [$10^{-4}/nm^2$] | 13.5 ± 1.5 |
| $\rho_{1,tail}$ [$10^{-4}/nm^2$] | 8.5 ± 1 |
| $\rho_{2,tail}$ [$10^{-4}/nm^2$] | 7.5 ± 1 |
| $\rho_{1,CH_3}$ [$10^{-4}/nm^2$] | 6.5 ± 1 |
| $d_{1,head}$ [nm] | 0.55 ± 0.05 |
| $d_{2,head}$ [nm] | 0.3 ± 0.1 |
| $D_{1,tail}$ [nm] | 2.5 ± 0.5 |
| $D_{2,tail}$ [nm] | 2.1 ± 0.5 |
| $d_{1,CH_3}$ [nm] | 0.4 ± 0.1 |
| $d_{w,2}$ [nm] | 0.5 ± 0.1 |
| $\rho_{3,head}$ [$10^{-4}/nm^2$] | 15 ± 2 |
| $\rho_{4,head}$ [$10^{-4}/nm^2$] | 14 ± 2 |
| $\rho_{3,tail}$ [$10^{-4}/nm^2$] | 8.2 ± 1 |
| $\rho_{4,tail}$ [$10^{-4}/nm^2$] | 8.5 ± 1 |
| $\rho_{2,CH_3}$ [$10^{-4}/nm^2$] | 5 ± 1 |
| $d_{3,head}$ [nm] | 0.4 ± 0.1 |
| $d_{4,head}$ [nm] | 0.55 ± 0.1 |
| $D_{3,tail}$ [nm] | 2.6 ± 0.2 |
| $D_{4,tail}$ [nm] | 2.3 ± 0.2 |
| $d_{2,CH_3}$ [nm] | 0.3 ± 0.1 |

Table S2: Values of structural parameters obtained from the fit of the NR data for a highly charged DPPS double bilayer.

| Parameter | T = 40°C |
|---|---|
| $d_{SiO_2}$ [nm] | 0.97 ± 0.1 |
| $SLD_{SiO_2}$ [$10^{-4}$/nm$^2$] | 3.47 |
| $f_{w,SiO_2}$ [%] | 0 |
| $\sigma_{SiO_2}$ [nm] | 0.4 ± 0.2 |
| $d_{w,1}$ [nm] | 1.1±0.2 |
| $SLD_{w,1}$ [$10^{-4}$/nm$^2$] | $SLD_{solvent}$ |
| $\sigma_{w,1}$ [nm] | 0.7 ± 0.2 |
| $d_{h11}$ [nm] | 0.7 ± 0.2 |
| $SLD_{h11}$ [$10^{-4}$/nm$^2$] | 2.63 |
| $f_{w, h11}$ [%] | 10± 2 |
| $\sigma_{h11/ch1}$ [nm] | 0.2 ± 0.1 |
| $d_{ch1}$ [nm] | 3.8 ±0.2 |
| $SLD_{ch1}$ [$10^{-4}$/nm$^2$] | -0.41 |
| $f_{w, ch1}$ [%] | 0 |
| $\sigma_{ch1/h12}$ [nm] | 0.2 ± 0.1 |
| $d_{h12}$ [nm] | 0.8 ± 0.2 |
| $SLD_{h12}$ [$10^{-4}$/nm$^2$] | 2.63 |
| $f_{w, h12}$ [%] | 10 ± 2 |
| $\sigma_{h12/w}$ [nm] | 0.4 ± 0.2 |
| $d_{w,2}$ [nm] | 0.6 ±0.2 |
| $SLD_{w,2}$ [$10^{-4}$/nm$^2$] | $SLD_{solvent}$ |
| $\sigma_{w,2}$ [nm] | 0.2 ± 0.1 |
| $d_{h21}$ [nm] | 0.8 ± 0.2 |
| $SLD_{h21}$ [$10^{-4}$/nm$^2$] | 2.63 |
| $f_{w, h21}$ [%] | 10±3 |
| $\sigma_{h21/w}$ [nm] | 0.2 ± 0.1 |
| $d_{ch2}$ [nm] | 3.8 ±0.2 |
| $SLD_{ch2}$ [$10^{-4}$/nm$^2$] | -0.41 |
| $f_{w, ch2}$ [%] | 0 |
| $\sigma_{ch2/h22}$ [nm] | 0.2 ± 0.1 |
| $d_{h22}$ [nm] | 0.8 ± 0.2 |
| $SLD_{h22}$ [$10^{-4}$/nm$^2$] | 2.63 |
| $f_{w, h22}$ [%] | 10±3 |
| $\sigma_{h22/w}$ [nm] | 0.4 ± 0.2 |

Table S3: Values of structural parameters obtained from the fit of the NR data for a highly charged DPPS triple bilayer.

| Parameter | T = 25°C |
|---|---|
| $d_{SiO_2}$ [nm] | 1.5 |
| $SLD_{SiO_2}$ [$10^{-4}$/nm$^2$] | 3.47 |
| $f_{w,SiO_2}$ [%] | 0 |
| $\sigma_{SiO_2}$ [nm] | 0.7 |
| $d_{w,1}$ [nm] | 1.3±0.2 |
| $SLD_{w,1}$ [$10^{-4}$/nm$^2$] | $SLD_{solvent}$ |
| $\sigma_{w/h11}$ [nm] | 0.7 ± 0.2 |
| $d_{h11}$ [nm] | 0.9 ± 0.2 |
| $SLD_{h11}$ [$10^{-4}$/nm$^2$] | 2.63 |
| $f_{w,h11}$ [%] | 25± 3 |
| $\sigma_{h11/ch1}$ [nm] | 0.3 ± 0.1 |
| $d_{ch1}$ [nm] | 3.5 ±0.2 |
| $SLD_{ch1}$ [$10^{-4}$/nm$^2$] | -0.41 |
| $f_{w, ch1}$ [%] | 0 |
| $\sigma_{ch1/h12}$ [nm] | 0.4 ± 0.2 |
| $d_{h12}$ [nm] | 0.9 ± 0.2 |
| $SLD_{h12}$ [$10^{-4}$/nm$^2$] | 2.63 |
| $f_{w, h12}$ [%] | 29 ± 3 |
| $\sigma_{h12/w}$ [nm] | 0.4 ± 0.2 |
| $d_{w,2}$ [nm] | 0.32 ±0.2 |
| $SLD_{w,2}$ [$10^{-4}$/nm$^2$] | $SLD_{solvent}$ |
| $\sigma_{w/h21}$ [nm] | 0.2 ± 0.1 |
| $d_{h21}$ [nm] | 0.9 ± 0.2 |
| $SLD_{h21}$ [$10^{-4}$/nm$^2$] | 2.63 |
| $f_{w, h21}$ [%] | 23±3 |
| $\sigma_{h21/ch2}$ [nm] | 0.4 ± 0.1 |
| $d_{ch2}$ [nm] | 3.8 ±0.2 |
| $SLD_{ch2}$ [$10^{-4}$/nm$^2$] | -0.41 |
| $f_{w, ch2}$ [%] | 0 |
| $\sigma_{ch2/h22}$ [nm] | 0.4 ± 0.2 |
| $d_{h22}$ [nm] | 0.9 ± 0.2 |
| $SLD_{h22}$ [$10^{-4}$/nm$^2$] | 2.63 |
| $f_{w, h22}$ [%] | 24±3 |
| $\sigma_{h22/w}$ [nm] | 0.4 ± 0.2 |
| $d_{w,3}$ [nm] | 0.31 ±0.2 |
| $\sigma_{w/h31}$ [nm] | 0.3 ± 0.1 |
| $d_{h31}$ [nm] | 0.9 ± 0.2 |
| $SLD_{h31}$ [$10^{-4}$/nm$^2$] | 2.63 |
| $f_{w,h31}$ [%] | 0.1 ± 0.02 |
| $\sigma_{h31/ch3}$ [Å] | 0.3 ± 0.1 |
| $d_{ch3}$ [nm] | 3.8 ±0.2 |
| $SLD_{ch3}$ [$10^{-4}$/nm$^2$] | -0.41 |
| $f_{w, ch3}$ [%] | 0 |
| $\sigma_{ch3/h32}$ [nm] | 0.2 ± 0.1 |
| $d_{h32}$ [nm] | 0.9 ±0.2 |
| $SLD_{h32}$ [$10^{-4}$/nm$^2$] | 2.63 |
| $f_{w,h32}$ [%] | 10±2 |
| $\sigma_{h32/w}$ [nm] | 0.3 ± 0.1 |

## 5   Osmotic pressure effect

We investigated the effect of ionic strength on the double bilayer system by adding NaCl up to a concentration of 0.3M. This corresponds to a Debye length $\ell_D \sim 0.8$ nm, of the order of the water spacing $d_{w,1}$ and $d_{w,2}$, not compatible with a screening effect. Most probably, as previously described by Hishida *et al.* [S10], the co-ions are expelled from the interbilayer spacing, resulting in an osmotic pressure $P_{osmo}$. As $d_{w,1} > d_{w,2}$, the interaction potential between the two bilayers is stiffer than that between the supported bilayer and the substrate. As a result, the osmotic effect only affects $d_{w,1}$. Assuming that the coions are expelled from the interlayer water spaces of thickness $d_{w,1}$ and $d_{w,2}$, an osmotic pressure $P_{osmo} = c k_B T$ will act on the system, where $c$ is the coions concentration, directly related to the Debye length by $c = \epsilon_w k_B T/(2e^2 \ell_D^2)$. The total potential for the first bilayer can be written as

$$U_{1,\text{salt}}(z) = U_{1,\text{no salt}}(z) + P_{osmo} z, \tag{S2}$$

where $U_{1,\text{no salt}}(z)$ is the interaction potential without salt. To avoid describing this potential in detail and to highlight only the effects of osmotic pressure, we can make a quadratic approximation of the potential :

$$U_{1,\text{salt}}(z) = \frac{1}{2} U''_{1,\text{no salt}} \left[z - d_{w,1}(P_{osmo} = 0)\right]^2 + P_{osmo} z, \tag{S3}$$

where $d_{w,1}(P = 0)$ is the equilibrium position without added salt. By minimizing equation S3, we obtain an approximate expression for the water thickness in presence of salt :

$$d_{w,1} = d_{w,1}(P = 0) - \frac{P}{U''} = d_{w,1}(P = 0) - \frac{k_B T}{8\pi U''_{1,\text{no salt}}} \frac{1}{\ell_D^2}. \tag{S4}$$

The expression S4 has been used to fit the data in Fig.4(a) (main paper) with only one fitting parameter $U''_{1,\text{no salt}}$. We found $U''_{1,\text{no salt}} \simeq 10^{11}$ J.m$^{-4}$ in good qualitative agreement with our previous work [S11].



\end{document}